%% file: apssamp.tex
\documentclass[%
 reprint,
nofootinbib,
 amsmath,amssymb,
 aps,
nolongbibliography
]{revtex4-2}

\usepackage{graphicx}

\usepackage{dcolumn}
\usepackage{bm}
\usepackage{braket} 

\usepackage{placeins}
\usepackage{xcolor}
\usepackage[colorlinks=true,
            linkcolor=blue,
            citecolor=blue,
            urlcolor=blue]{hyperref}

\begin{document}

\preprint{APS/123-QED}

\title{Extended Rydberg Lifetimes in a Cryogenic Atom Array}

\author{Junlan Jin${^{1,2,*}}$, Yue Shi${^{1,*}}$, Youssef Aziz Alaoui$^{1,\dagger}$, Jingxin Deng${^1}$, Yukai Lu${^1}$, Jeff D. Thompson${^2}$, Waseem S. Bakr${^{1,\ddagger}}$}

\affiliation{${^1}$Department of Physics, Princeton University, Princeton, New Jersey 08544, USA\\${^2}$Department of Electrical and Computer Engineering, Princeton University, Princeton, New Jersey 08544, USA}

\date{\today}

\begin{abstract}
We report on the realization of a $^{133}$Cs optical tweezer array in a cryogenic blackbody radiation (BBR) environment. By enclosing the array within a 4\,K radiation shield, we measure long Rydberg lifetimes, up to $406 (36)\,\mu$s for the $55 P_{3/2}$ Rydberg state, a factor of 3.3(3) longer than the room-temperature value. We employ single-photon coupling for coherent manipulation of the ground-Rydberg qubit. We measure a small differential dynamic polarizability of the transition, beneficial for reducing dephasing due to light intensity fluctuations. Our results pave the path for advancing neutral-atom two-qubit gate fidelities as their error budgets become increasingly dominated by $T_1$ relaxation of the ground-Rydberg qubit.
\end{abstract}

\maketitle
\def\thefootnote{*}\footnotetext{These authors contributed equally to this work}\def\thefootnote{\arabic{footnote}}
\def\thefootnote{$\dagger$}\footnotetext{Present address: Laboratoire Kastler Brossel, Coll\`{e}ge de France, CNRS, ENS-PSL University, Sorbonne Universit\'{e}, 11 Place Marcelin Berthelot, 75005 Paris, France}\def\thefootnote{\arabic{footnote}}
\def\thefootnote{$\ddagger$}\footnotetext{Email: wbakr@princeton.edu}\def\thefootnote{\arabic{footnote}}


Neutral atoms in optical tweezer arrays are a competitive platform for the realization of large-scale fault-tolerant quantum computers. Recent progress has included the demonstration of dynamical reconfigurability \cite{coherent-transport_bluvstein2022}, large-scale arrays containing thousands of qubits with long coherence times \cite{6100tweezer_manetsch2025},
continuous operation with atom reloading \cite{continuous_Chiu2025,continuous_Li2025}, and logical and mid-circuit operations \cite{logical_Bluvstein2024, atom-computing_Reichardt2025, atom-computing_Muniz2025, erasure_Zhang2025, fault-tolerant_Bluvstein2026}. In particular, two-qubit gate fidelities, the bottleneck for quantum error correction, have been improved beyond the surface code threshold for fault-tolerant quantum computation \cite{99.5-fidelity_Evered2023, 99.4_Michael2025,99.3_Radnaev2025, AtomComputing2025Gate, Aruku2025}, reaching up to $99.7\%$~\cite{99.7_Tsai2025}. Further increases in the gate fidelity can significantly reduce the overhead for quantum error correction~\cite{roffe2019}.

While qubits are usually encoded in a pair of long-lived hyperfine ground states, two-qubit gates rely on coupling a ground state to a Rydberg state. 
Focusing on the ground-Rydberg subspace, errors are primarily due to longitudinal and transverse relaxation during the gate operation (characterized by $T_1$ and $T^*_2$, respectively) and off-resonant couplings to Rydberg Zeeman sublevels outside the subspace \cite{99.5-fidelity_Evered2023}. As gate fidelities have improved, $T_1$ relaxation is rapidly becoming the dominant error source.
It is determined by spontaneous emission from the Rydberg state, BBR-induced transitions to nearby Rydberg states and intermediate state scattering in commonly used two-photon excitation schemes \cite{99.3_Radnaev2025, 99.5-fidelity_Evered2023}.
In state-of-the-art optical tweezer array experiments, the relaxation time $T_1$ is typically
below $65\,\mu$s
\cite{99.5-fidelity_Evered2023, 99.7_Tsai2025, Aruku2025, 99.4_Michael2025}. A promising route to extend $T_1$ is the suppression of BBR-induced transitions using a cryogenic environment~\cite{Gallagher1979BBRTheory, Spencer1981Na, Spencer1982Na}.

Previous experiments have prepared atom arrays inside low-temperature cryostats, demonstrating improved trapping lifetimes of the atoms as well as excitation to Rydberg states \cite{6000s_Schymik2021, 2000rearrange_Pichard2024, 3000s_Zhang2025}.
Long vacuum lifetimes are advantageous for assembling large defect-free arrays and increasing imaging fidelity for mid-circuit detection. 
However, a demonstration of the other advantage of cryogenic tweezer array experiments, the extended Rydberg lifetimes due to suppressed BBR-induced transitions, has remained outstanding. 

In this work, we demonstrate the extension of Rydberg state lifetimes for cesium atom arrays prepared in a cryogenic blackbody radiation environment.
For the state $55P_{3/2}$, we measure a lifetime of $406(36)\,\mu$s, corresponding to an effective BBR temperature of less than $25\,$K at the 1$\sigma$ uncertainty level. In this temperature range, BBR-induced transitions are suppressed by more than one order of magnitude compared to room temperature. We also demonstrate coherent manipulation of the ground-Rydberg qubit using single-photon coupling, which avoids intermediate state scattering present in two-photon coupling schemes. Together, these features of our experiment push $T_1$ towards the fundamental limit set by spontaneous emission of the Rydberg state.


\begin{figure*}[t]
  \centering
  \includegraphics[width=\textwidth]{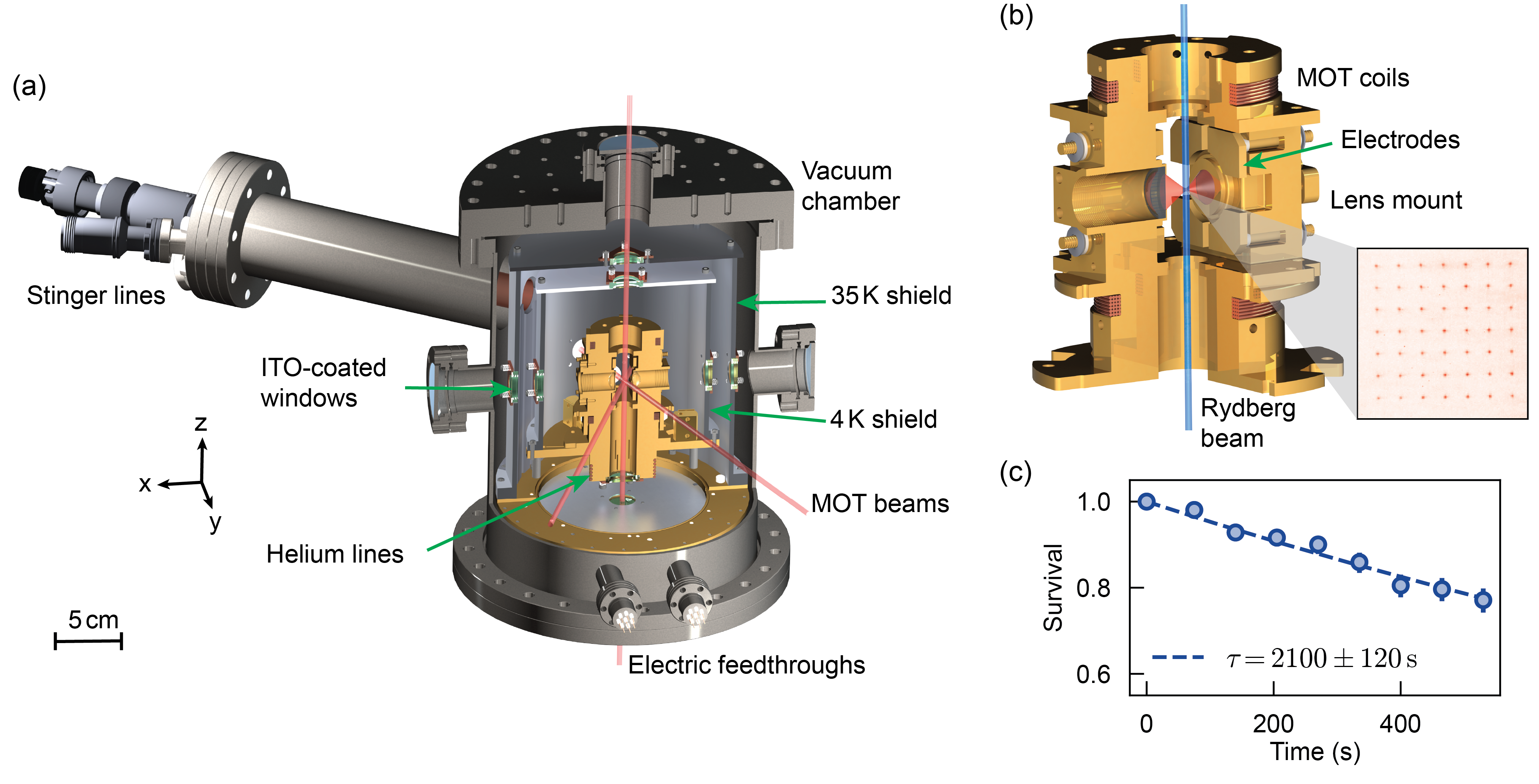}
  \caption{ (a) UHV cryostat with 4K and 35K radiation shields inside a room-temperature vacuum chamber. The atomic beam (not shown) is directed along the $y$ axis. (b) Assembly inside the 4K radiation shield, integrating aspheric lenses, electrodes and MOT coils. Inset: averaged fluorescence image of a $7\times7$ atom array, with a tweezer spacing of 12\,$\mu$m. (c) Vacuum trapping lifetime measurement. 
  }
  \label{fig1:setup}
\end{figure*}

Our apparatus consists of two vacuum chambers connected by a differential pumping tube: a room-temperature atom-source chamber and a cryogenic science chamber [see Fig.\,\ref{fig1:setup}\,(a)]. A beam of cesium-133 atoms is produced from a 2D magneto-optical trap (MOT) in the room-temperature chamber, and delivered to the ultra-high vacuum (UHV) cryostat using a push beam. There, the atoms are trapped in a 3D MOT, and stochastically loaded into an optical tweezer array. The core assembly features a pair of aspheric lenses with a numerical aperture of 0.5 for projecting optical tweezers and imaging the atom array, superconducting coils for producing the magnetic field for the MOT, and electrodes for controlling the electric field environment [see Fig.\,\ref{fig1:setup}\,(b)]. This assembly is mounted on a 4\,K baseplate. Helium gas flows through copper tubing brazed on the baseplate, providing a cooling power of 0.4\,W at the base temperature. We use a closed-cycle system (ColdEdge Technologies, Stinger), where the helium gas circulates between a Gifford-McMahon cryocooler and the cryostat, with the flexible helium line damping mechanical vibrations at the baseplate. 

To ensure a low-BBR environment, two radiation shields surround the assembly, at 35\,K and 4\,K, respectively. Both shields have windows for laser beams and a small opening for the atomic beam. Blackbody radiation from the room-temperature walls of the cryostat chamber can leak through cold windows on the shields, increasing the effective BBR temperature at the location of the atoms. To address this issue, we coat the windows on the 4\,K shield with a 120\,nm thick layer of indium tin oxide (ITO) with sheet resistance $R_{\square}\sim 20\,\Omega$/sq, which transmits optical beams (transmission coefficient $T\approx95\%$) while suppressing microwaves in the $10-300$\,GHz frequency range, relevant for transitions between Rydberg states ($T\lesssim3\%$) ~\cite{ITO_Meinert2020}. For the UV-beam path for Rydberg excitation, we leave the central region of the windows uncoated since the ITO coating has a strong absorption of UV light.
For the tweezer-beam path which has a higher laser power, we use a 30\,nm thick ITO coating ($R_{\square}\sim 100\,\Omega$/sq) and put the coating on the 35\,K windows since the cooling power at that temperature is higher, and can handle the 1\% laser power absorbed per window from the ITO coating. 

Electric field control is crucial for manipulating Rydberg atoms, especially for $nP$ Rydberg states used in this experiment, which have a larger dc polarizability than $nS$ Rydberg states accessible with two-photon excitation. For example, the polarizability of the $\ket{55P_{3/2},m_J = \pm 1/2}$ state is $2740\,$MHz\,cm$^2$/V$^2$, a factor of 26 larger than the $\ket{55S_{1/2},m_J = \pm 1/2}$ state \cite{arc_sibalic2017}. We use eight electrodes connected to vacuum feedthroughs using twisted-pair cables. These cables allow the transmission of dc voltages for compensating stray electric fields, as well as RF signals with frequencies up to a few hundred MHz, while strongly attenuating GHz thermal noise from the 300\,K environment that can drive BBR-induced transitions.
The wires are heat sunk at both 4\,K and 35\,K stages. Although RF signals are not used in the current experiment, this design adds flexibility for future experiments with circular Rydberg atoms, where RF drives with well-defined polarization are needed for transfer to circular states. Compared to the $nP$  Rydberg states used in this work, atoms in circular states can exhibit even more dramatic lifetime enhancements at cryogenic temperatures \cite{circularRydberg_Nguyen2018, cantat2020long, circularRydberg_Cohen2021}. 

We start our experiments by loading atoms from the 3D MOT into a tweezer array [see Fig.\,1(b) inset]. Then, we apply two stages of polarization gradient cooling (PGC), between which we adiabatically ramp down the tweezer trap depth and image the initial tweezer occupancies (see Supplemental Material~\cite{supp}).

We first characterize the improvement in vacuum-limited atom lifetimes. An ion gauge located at the room-temperature region of the cryostat reads $2\times10^{-10}$ Torr after cooldown. However, the 4\,K and 35\,K surfaces provide a strong cryopumping effect, further reducing the pressure inside the radiation shields.
During the lifetime measurement, we hold the atoms at a depth of 210\,$\mu$K and apply pulsed PGC with a  detuning of $-23\,\Gamma$ for 10\,ms every 2\,s \cite{6000s_Schymik2021,6100tweezer_manetsch2025}, where $\Gamma = 2\pi\times5.22\,$MHz is the natural linewidth of the 
$\ket{6S_{1/2}, F=4}\rightarrow \ket{6P_{3/2}, F'=5}$ transition. Atoms are imaged every 65\,s.
A $1/e$ lifetime of 2100\,s is extracted in Fig.\,\ref{fig1:setup}\,(c) after accounting for the $5\times10^{-3}$ loss per imaging step, obtained from a separate measurement. We note that this lifetime is obtained with an unbaked vacuum chamber, only utilizing a 150~L/s ion pump in combination with cryopumping. Further improvements may be achievable by baking the system and firing an integrated titanium sublimation pump. 


\begin{figure}[t]
  \centering
  \includegraphics[width=1\linewidth]{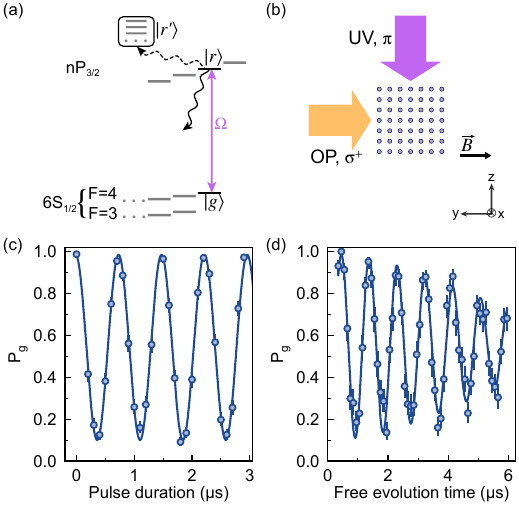}
  \caption{
  Coherent control of the ground-Rydberg qubit.
  (a) Relevant energy levels and decay pathways. Atoms are initialized into $\ket{g} = \ket{6S_{1/2}, F=4, m_F= 4}$ via optical pumping and excited to the Rydberg state $\ket{r} = \ket{nP_{3/2}, m_J =1/2}$ using single-photon excitation at 319\,nm. The Rydberg state can undergo BBR-induced transitions to other Rydberg states $\ket{r'}$ or spontaneous decay to low-lying states, indicated by curvy arrows.
  (b) Configuration of the UV excitation and optical pumping (OP) beams relative to the atom array. The UV and OP beams have $\pi$ and $\sigma^+$ polarizations, respectively. The quantization axis is defined by the magnetic field $B$ along the $y$ axis.
 (c) Rabi oscillation between $\ket{g}$ and $\ket{r}$ for the $55P_{3/2}$ Rydberg state, averaged over atoms in the central column along the $z$ direction. $P_g$ denotes the ground-state fraction.
 (d) Ramsey measurement for a single atom at the center of the array, yielding a fitted coherence time $T_2^*$ = 6.2(4)$\mu$s.
  }
  \label{fig2:Rydberg_control}
\end{figure}

Next, we demonstrate coherent manipulation of the ground-Rydberg qubit. We employ single-photon excitation to $nP$ Rydberg states, avoiding the reduction of $T_1$ associated with intermediate state scattering in two-photon excitation schemes. The excitation light at 319\,nm is generated via sum-frequency generation of a 1560.5\,nm fiber laser and a 1080\,nm external-cavity diode laser (ECDL), followed by second-harmonic generation, producing up to 300\,mW out of a solarization-resistant UV fiber \cite{uvfiber_colombe2014}. The UV frequency can be tuned over a wide range of Rydberg states by adjusting the ECDL frequency. 

The relevant energy levels and beam configuration for Rydberg excitation are shown in Fig.\,\ref{fig2:Rydberg_control}\,(a)-(b). After the two stages of PGC, we optically pump the atoms into the stretched state $\ket{g} = \ket{6S_{1/2}, F=4, m_F= 4}$ at a reduced trap depth of $50\,\mu$K, with a 3\,G bias magnetic field defining the quantization axis along the $y$ direction. We then lower the trap depth adiabatically to further cool the atoms. Subsequently, we turn off the optical tweezer light, use the Rydberg beam to drive the transition between $\ket{g}$ and $\ket{r} = \ket{nP_{3/2}, m_J =1/2}$, and then turn the tweezer light back on for detection. The Rydberg beam propagates along the vertical $z$ direction perpendicular to the quantization axis, and is focused to a waist ($1/e^2$ radius) of 80\,$\mu$m at the position of atoms. 
We initially observed a strong spatial gradient of the transition frequencies from background electric fields. By applying compensating fields with the electrodes, we suppressed the frequency variation across the array to less than 100\,kHz for $n=55$, corresponding to a field variation of 9\,mV/cm.

\begin{figure}[t]
  \centering
  \includegraphics[width=1\linewidth]{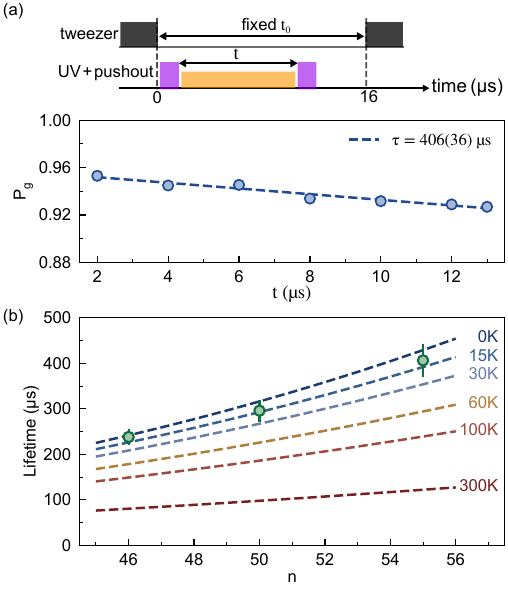}
  \caption{
  Extended $T_1$ lifetimes.
  (a) Experimental sequence and lifetime measurement of the $55P_{3/2}$ Rydberg state. $P_g(t)$, with $t$ being the gap time between two UV $\pi$ pulses, is fitted to an exponential decay with no offset. The fitted $1/e$ lifetime is 406(36)\,$\mu$s.
(b) Lifetimes of different $nP_{3/2}$ Rydberg states at different BBR temperatures. 
Green circles with error bars ($1\sigma$) denote experimental values measured for $n=46, 50, 55$. Dashed lines serve as guides to the eye connecting results calculated using ARC \cite{arc_sibalic2017}. 
  }
  \label{fig3:Lifetime}
\end{figure}

Figure~\ref{fig2:Rydberg_control}\,(c) shows Rabi oscillations between $\ket{g}$ and $\ket{r} = \ket{55P_{3/2}, m_J =1/2}$ at a Rabi frequency $\Omega = 2\pi \times 1.35\,$MHz. We detect ground-state atoms, while Rydberg atoms are ejected by the ponderomotive potential of the tweezer, with the contrast of the oscillation determined by imperfect ejection due to spontaneous Rydberg decay during the process. We also use a Ramsey sequence to characterize the coherence of the ground-Rydberg qubit [Fig.\,\ref{fig2:Rydberg_control}\,(d)]. The Gaussian decay of the envelope of the oscillations gives a Doppler-limited coherence time $T_2^* = 6.2(4)\,\mu$s, corresponding to a temperature of 2.2(3)\,$\mu$K.

Having established control over the ground-Rydberg qubit, we proceed to measure the extension of the Rydberg state lifetime at cryogenic temperatures using the protocol illustrated in Fig.~\ref{fig3:Lifetime}(a). The tweezer light is extinguished for a fixed duration of $t_0=16$\,$\mu$s, chosen such that in the absence of any light pulses, the atoms are recaptured in the tweezers with almost unit probability.
During the tweezer off time, a $\pi$ pulse is applied to transfer the population from $\ket{g}$ to $\ket{r}$ and after a variable delay $t$, a second $\pi$ pulse is applied to transfer the population back to $\ket{g}$. 
The transfer efficiency of each Rydberg $\pi$ pulse is 98-99\%, limited by residual intensity fluctuations of the Rydberg light and its inhomogeneity across the array. In between the two $\pi$ pulses, we apply a strong pushout beam to continuously blow away any population in the $6S_{1/2}$ manifold from decay of the Rydberg state. The pushout beam is resonant with the $\ket{6S_{1/2}, F=4}\rightarrow \ket{6P_{3/2}, F'=5}$ transition and a repump beam resonant with $\ket{6S_{1/2}, F=3}\rightarrow \ket{6P_{3/2}, F'=4}$ is applied simultaneously (see also Supplemental Material~\cite{supp}). 
A lifetime measurement curve for $\ket{55P_{3/2}, m_J =1/2}$ is shown in Fig.\,\ref{fig3:Lifetime}\,(a). Each data point is averaged over about 2400 repeated measurements.
The measured lifetime, obtained from an exponential decay fit, is 406(36)\,$\mu$s. This is a factor of 3.3(3) longer than the room-temperature lifetime of 122\,$\mu$s and approaches the spontaneous-decay-limited lifetime of 429\,$\mu$s, both calculated using ARC \cite{arc_sibalic2017}. Comparison with ARC calculations indicates that this lifetime corresponds to an effective BBR temperature of $10^{+13}_{-10}$\,K, indicating a strong suppression of BBR-induced transitions. At 10\,K, we calculate that the BBR-induced decay rate of the 55$P_{3/2}$ state is suppressed by a factor of 40 compared to room temperature.

Our measured  lifetimes for $n = 55, 50, 46$, shown in Fig.\,\ref{fig3:Lifetime}\,(b), are all about three times longer than the room-temperature values and consistent with an effective BBR temperature of $\lesssim 25\,$K. Further reduction of the BBR temperature would only lead to marginal lifetime improvement, as it is primarily limited by spontaneous decay. Even longer lifetimes are expected for higher $n$ states, given the $n^3$ scaling of the radiative lifetime. However, we restrict our attention to states in this range of $n$ because of their utility for implementing two-qubit gates in future work. The $n^7$ scaling of the dc polarizability of Rydberg states rapidly degrades $T^*_2$ at higher $n$ due to coupling to stray electric fields.

The extended Rydberg lifetime is expected to directly translate into improved two-qubit gate fidelity. In state-of-the-art experiments, the $T_1$ contribution to two-qubit gate infidelity is at the $(1-3)\times10^{-3}$ level~\cite{99.5-fidelity_Evered2023,99.7_Tsai2025,99.4_Michael2025} and is becoming the dominant error source. Because this infidelity scales as $1/(\Omega T_1)$, extending the lifetime reduces the gate error for any given drive strength. Even with the modest Rabi frequency $\Omega=2\pi\times1.35$\,MHz used here, we project the $T_1$ contribution to infidelity to be $6.4\times 10^{-4}$ for a time-optimal gate \cite{time-optimal_Jandura2022, NMR_grape_Khaneja2005}, already below the typical floor of room-temperature experiments. By reducing the beam waist to $20\,\mu$m to increase the Rabi frequency, a typical size balancing drive strength against intensity noise \cite{99.5-fidelity_Evered2023,99.7_Tsai2025, 99.4_Michael2025}, this contribution would be further suppressed to $1.6\times10^{-4}$. In the context of quantum error correction \cite{qec_Shor1995,css_code_Gottesman1997,qec_Knill1997}, such a reduction in physical error rates of two-qubit gates leads to strong suppression of the logical error rates. For instance, considering physical two-qubit gates limited by $T_1$ and a moderate size surface code with distance $d=7$ \cite{logical_Bluvstein2024}, our improvement of the Rydberg lifetime by a factor of 3.3 relative to room temperature would lead to an approximately two-orders-of-magnitude reduction in the logical error rate \cite{Surface_code_Fowler2012}.


\begin{figure}[t]
  \centering
  \includegraphics[width=0.95\linewidth]{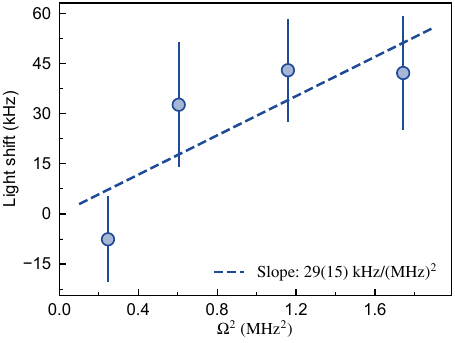}
  \caption{
  Light shift of the $55P_{3/2}$ Rydberg transition as a function of the square of the Rabi frequency $\Omega^2$. A linear fit (dashed line) to the data gives a light-shift coefficient $\kappa = 29(15)\,$kHz/MHz$^2$.
  }
  \label{fig4:light_shift}
\end{figure}

Achieving high Rabi frequencies to suppress the contribution of $T_1$ to gate errors requires management of light shifts associated with Rydberg excitation light. In two-photon excitation schemes used in alkali atom quantum processors, off-resonant coupling to the intermediate states induces strong differential light shifts between the ground and Rydberg states~\cite{99.5-fidelity_Evered2023}. Therefore, temporal noise and spatial inhomogeneities in the Rydberg beam intensity are converted into effective frequency noise, introducing a dephasing channel that grows with Rabi frequency and limits gate fidelity. We verify that our use of single-photon excitation significantly mitigates this issue. Performing spectroscopy of the Rydberg transition frequency at different Rabi coupling strengths, we measure a light-shift coefficient of $\kappa = 29(15)\,$kHz/MHz$^2$ (Fig.\,\ref{fig4:light_shift}).
Similar to observations with alkaline earth qubits \cite{madjarov2020,erasure_Zhang2025}, the light shift is negligibly small at our current Rabi frequency and should remain well below the Rabi frequency even when reducing the beam waist as described above. 

Beyond the absolute extension of $T_1$, the specific suppression of BBR-induced transitions is beneficial for reducing correlated errors in quantum circuits. Compared to spontaneous emission, BBR-induced transitions can be more detrimental due to collective effects. Consider, for example, a transversal entangling gate between logical qubits, as demonstrated in~\cite{logical_Bluvstein2024}. A BBR transition transfers an atom to a different Rydberg state that couples to the rest of the excited atoms via long-range dipole-dipole interactions, inducing correlated dephasing and loss~\cite{Broadening_Goldschmidt2016,avalanche_Boulier2017,Cong2022,fault-tolerant_Bluvstein2026}. In fact, even at our cryogenic temperatures, we have observed collective effects in our lifetime measurements. To extract single atom Rydberg lifetimes, we use a $3\times3$ tweezer array with 24\,$\mu$m spacing to perform the measurements. In contrast, for a $7\times7$ tweezer array with 12\,$\mu$m spacing, we observe accelerated and non-exponential decay of the 
probability to be in the ground state
at the end of the lifetime measurement protocol (see Supplemental Material~\cite{supp}).


In summary, we have realized a cryogenic neutral atom platform engineered to suppress blackbody radiation. Using this platform, we demonstrated a substantial extension of Rydberg state lifetimes, reducing BBR-induced transitions by more than an order of magnitude. The extension of Rydberg lifetimes provides a pathway for further improvements of two-qubit gate fidelities in neutral atom quantum computers by reducing the dominant source of incoherent errors. Furthermore, our platform opens new possibilities for other Rydberg-based quantum applications. For instance, the low-BBR environment can reduce collective avalanche loss that currently limits Rydberg-dressing schemes for quantum simulation and metrology~\cite{Dressing_Zeiher2016, Schleier-Smith2023DressingSqueezing, Cao2024DressingMetrology}. It also enables extended lifetimes for circular Rydberg states~\cite{cantat2020long}, a promising resource for quantum simulation and computation applications~\cite{circularRydberg_Nguyen2018,circularRydberg_Cohen2021}. In contrast to recent approaches utilizing room-temperature parallel-plate capacitors \cite{circular-Rydberg_Wu2023,Circular_Sr_Holzl2024,circularRydberg_Pultinevicius2025}, our platform provides a broadband solution applicable across a wide range of Rydberg levels.\\

\textit{Acknowledgements}--- We would like to thank Ourania-Maria Glezakou-Elbert for experimental assistance, Chenyuan Li for help with UV fibers, and Donghyuk Seo for calculations of gate fidelities.  We also thank Lawrence Cheuk's group and Antoine Browaeys for helpful discussions. We thank Max Prichard for feedback on the manuscript.
This work was supported by the the National Science Foundation (QLCI grant OMA-2120757), the Brown Science Foundation and the David and Lucile Packard Foundation (grant no. 2016-65128).

\textit{Competing interests}--- J.D.T is a shareholder in Logiqal, Inc.

\bibliography{apssamp}

\clearpage
\onecolumngrid

\section*{Supplemental Material}

\setcounter{section}{0}
\setcounter{figure}{0}
\setcounter{table}{0}
\setcounter{equation}{0}

\renewcommand{\thesection}{S\arabic{section}}
\renewcommand{\thefigure}{S\arabic{figure}}
\renewcommand{\thetable}{S\arabic{table}}
\renewcommand{\theequation}{S\arabic{equation}}

\input{a_supp_body}

\end{document}

%% file: a_supp_body.tex
\section{Experimental sequence}
\label{appedix:sequence}

The 3D MOT beams consist of two pairs of counter propagating horizontal  cooling beams with 1.6\,mm waist ($1/e^2$ radius), and a retroreflected vertical cooling beam with 3\,mm waist. For each cooling beam, there is a co-propagating repump beam. The cooling beams have a detuning of $\Delta = -3\,\Gamma$ relative to the bare  $\ket{6S_{1/2}, F=4}\rightarrow \ket{6P_{3/2}, F'=5}$ atomic transition frequency, where $\Gamma = 2\pi\times5.22\,$MHz is the natural linewidth of this transition. The repump beam is at the resonance of the bare $\ket{6S_{1/2}, F=3}\rightarrow \ket{6P_{3/2}, F'=4}$ transition. The magnetic field of the MOT is generated by a pair of superconducting coils inside the 4K radiation shield. We apply a current of 1.2\,A through both coils, producing a magnetic field gradient of 8.4\,G/cm.

Atoms are first loaded from a 3D MOT into optical tweezers with 1.15\,$\mu$m waist, 700\,$\mu$K trap depth and 57\,kHz radial trap frequency. Using a loading time of 600\,ms, we achieve a typical array filling fraction of $\sim0.5$. The tweezer light is at 936\,nm, a magic wavelength for the cesium D2 line.
Following loading, we apply the first stage of polarization gradient cooling (PGC) by increasing the detuning of the cooling beams to $\Delta=-17\,\Gamma$ and reducing the cooling beam power. The tweezer trap depth is then ramped down to $210\,\mu\text{K}$, where atoms are imaged with a $60\,\text{ms}$ exposure time at $\Delta = -15\,\Gamma$. Subsequently, we apply the second stage of PGC with $\Delta=-23\,\Gamma$. After lowering the trap depth to 50\,$\mu$K, we optically pump atoms for 3\,ms using a $\sigma^+$-polarized beam resonant with the $\ket{6S_{1/2}, F=4}\rightarrow \ket{6P_{3/2}, F'=4}$ transition, alongside a co-propagating $\sigma^+$-polarized repump beam. A $3\,\text{G}$ bias field along the $y$-axis defines the quantization axis. Finally, the trap depth is adiabatically lowered to $16\,\mu\text{K}$. The resulting state-preparation infidelity is $< 0.5\%$.

After adiabatic cooling, we turn off the tweezer trap to apply the Rydberg pulses for Rydberg excitation and $T_1$ lifetime measurements.

\section{Lifetime measurements}
\label{appedix:release-recap_push}

\begin{figure}[b]
  \centering
  \includegraphics[width=0.9\linewidth]{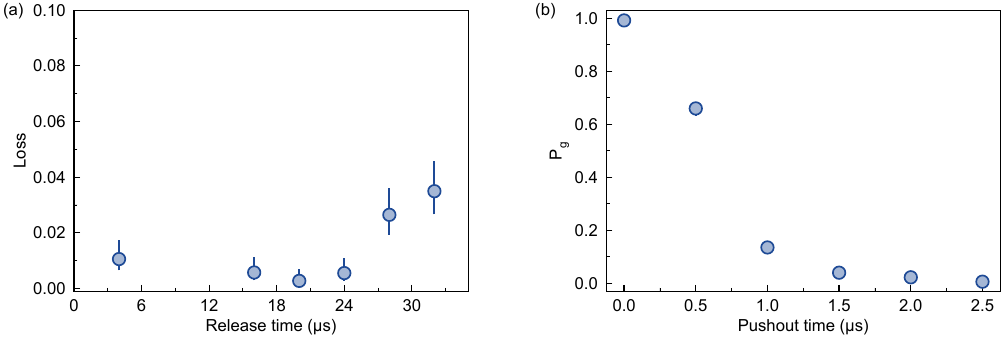}
  \caption{
  (a) Atom loss after different release times. When the release time is less than 24\,$\mu$s, there is essentially no loss during the release-recapture sequence. The approximately $0.5\%$ loss observed is consistent with imaging loss.
    (b) Pushout of atoms in the ground states. The atoms are not optically pumped into the stretched state before the pushout. The experimental protocol used is the same as that shown in Fig.\,3\,(a) of the main text, except that the Rydberg beam remains off.
  }
  \label{fig5:release_recap_push}
\end{figure}

We first measure the atom loss after different release times, without applying Rydberg pulses.
The loss remains negligible for release times up to 24\,$\mu$s as shown in Fig.\,\ref{fig5:release_recap_push}\,(a). The observed loss of about 0.5\% is consistent with the imaging loss.  We use a fixed release time of 16\,$\mu$s in the lifetime measurements to provide a conservative safety margin, ensuring recapture in the presence of effects on the atomic motion induced by UV photons.

We measure the lifetime by first exciting all atoms to the Rydberg state using a $\pi$ pulse. Then the Rydberg state undergoes spontaneous decay to the ground state or BBR-induced transitions to other Rydberg states. To determine the population remaining in the original Rydberg state after a certain amount of time, we apply a second $\pi$ pulse to transfer the population back to the ground state for detection. Between the two $\pi$ pulses, a strong pushout beam is continuously applied to blow out any atoms that decay to ground states. We characterize the efficiency of the pushout beam by removing the Rydberg pulses and optical pumping process to verify that all ground state hyperfine levels can be efficiently pushed out. A pushout curve measuring the recaptured population versus the pushout time is shown in Fig.\,\ref{fig5:release_recap_push}\,(b). 
In the Rydberg lifetime measurements [Fig.\,3 of the main text], the data are taken with gap times $t\geq2\,\mu$s between two Rydberg $\pi$ pulses, and with the pushout pulse duration set to about $t-1\,\mu$s.

We comment on the fact that an atom in the Rydberg state can be falsely detected as a ground state with nearly 10\% probability as seen from the lowest value in the Rabi oscillations in the main text. This false detection has negligible contribution to our measurement since atoms in the target Rydberg state are transferred to the ground state for detection at the end of the protocol.

\section{Collective effects}
\label{appedix:collective}
\begin{figure}[t]
  \centering
  \includegraphics[width=0.5\linewidth]{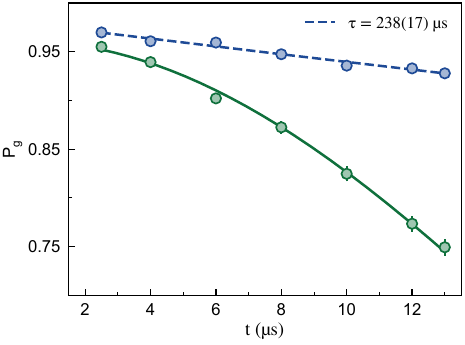}
  \caption{
  Ground-state fraction $P_g$ versus the gap time between two Rydberg $\pi$ pulses for the $\ket{46P_{3/2},m_J=1/2}$ state. Blue and green data points correspond to a $3\times3$ tweezer array with 24\,$\mu$m spacing and a $7\times7$ array with 12\,$\mu$m spacing, respectively. The dashed line shows an exponential fit to the blue data. The fitted lifetime is 238(17)\,$\mu$s. The green solid line is a guide to the eye.
  }
  \label{fig6:density}
\end{figure}

For the lifetime measurements, we operate with a reduced density array, using a $3\times3$ tweezer array with 24\,$\mu$m spacing.
When using a $7\times7$ array with 12\,$\mu$m spacing, the ground-state fraction decreases significantly faster and cannot be described by an exponential fit (Fig.\,\ref{fig6:density}).
We attribute this to collective effects induced by the transition of one Rydberg atom into a different Rydberg state. The resulting dipolar interaction perturbs the energies of nearby Rydberg atoms, reducing the fidelity of the second $\pi$ pulse and thereby lowering the transfer efficiency back to the ground state. For instance, the dipolar interaction between an atom in $\ket{46 P_{3/2},m_J=1/2}$ and an atom in a Rydberg state of opposite parity can reach up to 1\,MHz and 0.35\,MHz for nearest and next-nearest neighbors, respectively. Such interaction strengths are non-negligible compared to the Rabi frequency of the Rydberg $\pi$ pulses.
A more detailed investigation of the collective effects is left for future work.

%% file: apssamp.bib
@article{roffe2019,
  title={Quantum error correction: an introductory guide},
  author={Roffe, Joschka},
  journal={Contemp. Phys.},
  volume={60},
  number={3},
  pages={226--245},
  year={2019},
  publisher={Taylor \& Francis},
  url = {https://doi.org/10.1080/00107514.2019.1667078}
}

@article{madjarov2020,
	title = {High-fidelity entanglement and detection of alkaline-earth {Rydberg} atoms},
	volume = {16},
	copyright = {2020 The Author(s), under exclusive licence to Springer Nature Limited},
	issn = {1745-2481},
	url = {https://www.nature.com/articles/s41567-020-0903-z},
	doi = {10.1038/s41567-020-0903-z},
	language = {en},
	number = {8},
	urldate = {2026-02-02},
	journal = {Nat. Phys.},
	publisher = {Nature Publishing Group},
	author = {Madjarov, Ivaylo S. and Covey, Jacob P. and Shaw, Adam L. and Choi, Joonhee and Kale, Anant and Cooper, Alexandre and Pichler, Hannes and Schkolnik, Vladimir and Williams, Jason R. and Endres, Manuel},
	month = aug,
	year = {2020},
	pages = {857--861},
}

@article{uvfiber_colombe2014,
author = {Yves Colombe and Daniel H. Slichter and Andrew C. Wilson and Dietrich Leibfried and David J. Wineland},
journal = {Opt. Express},
number = {16},
pages = {19783--19793},
publisher = {Optica Publishing Group},
title = {Single-mode optical fiber for high-power, low-loss UV transmission},
volume = {22},
month = {Aug},
year = {2014},
url = {https://opg.optica.org/oe/abstract.cfm?URI=oe-22-16-19783},
doi = {10.1364/OE.22.019783}
}

@article{NMR_grape_Khaneja2005,
	title = {Optimal control of coupled spin dynamics: design of {NMR} pulse sequences by gradient ascent algorithms},
	volume = {172},
	issn = {1090-7807},
	shorttitle = {Optimal control of coupled spin dynamics},
	url = {https://www.sciencedirect.com/science/article/pii/S1090780704003696},
	doi = {10.1016/j.jmr.2004.11.004},
	number = {2},
	urldate = {2026-01-31},
	journal = {Journal of Magnetic Resonance},
	author = {Khaneja, Navin and Reiss, Timo and Kehlet, Cindie and Schulte-Herbrüggen, Thomas and Glaser, Steffen J.},
	month = feb,
	year = {2005},
	pages = {296--305},
}

@article{time-optimal_Jandura2022,
	title = {Time-{Optimal} {Two}- and {Three}-{Qubit} {Gates} for {Rydberg} {Atoms}},
	volume = {6},
	url = {https://quantum-journal.org/papers/q-2022-05-13-712/},
	doi = {10.22331/q-2022-05-13-712},
	language = {en-GB},
	urldate = {2026-01-30},
	journal = {Quantum},
	publisher = {Verein zur Förderung des Open Access Publizierens in den Quantenwissenschaften},
	author = {Jandura, Sven and Pupillo, Guido},
	month = may,
	year = {2022},
	pages = {712},
}

@misc{supp,
  note = {See Supplemental Material.}
}

@article{Aruku2025,
      author={Aruku Senoo and Alexander Baumgärtner and Joanna W. Lis and Gaurav M. Vaidya and Zhongda Zeng and Giuliano Giudici and Hannes Pichler and Adam M. Kaufman},
      year={2025},
      journal={arXiv:2506.13632},
      url={https://arxiv.org/abs/2506.13632}
}

@article{atom-computing_Reichardt2025,
      author={Ben W. Reichardt and Adam Paetznick and David Aasen and Ivan Basov and Juan M. Bello-Rivas and Parsa Bonderson and Rui Chao and Wim van Dam and Matthew B. Hastings and Ryan V. Mishmash and Andres Paz and Marcus P. da Silva and Aarthi Sundaram and Krysta M. Svore and Alexander Vaschillo and Zhenghan Wang and Matt Zanner and William B. Cairncross and Cheng-An Chen and Daniel Crow and Hyosub Kim and Jonathan M. Kindem and Jonathan King and Michael McDonald and Matthew A. Norcia and Albert Ryou and Mark Stone and Laura Wadleigh and Katrina Barnes and Peter Battaglino and Thomas C. Bohdanowicz and Graham Booth and Andrew Brown and Mark O. Brown and Kayleigh Cassella and Robin Coxe and Jeffrey M. Epstein and Max Feldkamp and Christopher Griger and Eli Halperin and Andre Heinz and Frederic Hummel and Matthew Jaffe and Antonia M. W. Jones and Eliot Kapit and Krish Kotru and Joseph Lauigan and Ming Li and Jan Marjanovic and Eli Megidish and Matthew Meredith and Ryan Morshead and Juan A. Muniz and Sandeep Narayanaswami and Ciro Nishiguchi and Timothy Paule and Kelly A. Pawlak and Kristen L. Pudenz and David Rodríguez Pérez and Jon Simon and Aaron Smull and Daniel Stack and Miroslav Urbanek and René J. M. van de Veerdonk and Zachary Vendeiro and Robert T. Weverka and Thomas Wilkason and Tsung-Yao Wu and Xin Xie and Evan Zalys-Geller and Xiaogang Zhang and Benjamin J. Bloom},
      year={2025},
      journal={arXiv:2411.11822},
      url={https://arxiv.org/abs/2411.11822}, 
}

@article{atom-computing_Muniz2025,
  title = {Repeated Ancilla Reuse for Logical Computation on a Neutral Atom Quantum Computer},
  author = {Muniz, J. A. and Crow, D. and Kim, H. and Kindem, J. M. and Cairncross, W. B. and Ryou, A. and Bohdanowicz, T. C. and Chen, C.-A. and Ji, Y. and Jones, A. M. W. and Megidish, E. and Nishiguchi, C. and Urbanek, M. and Wadleigh, L. and Wilkason, T. and Aasen, D. and Barnes, K. and Bello-Rivas, J. M. and Bloomfield, I. and Booth, G. and Brown, A. and Brown, M. O. and Cassella, K. and Cowan, G. and Epstein, J. and Feldkamp, M. and Griger, C. and Hassan, Y. and Heinz, A. and Halperin, E. and Hofler, T. and Hummel, F. and Jaffe, M. and Kapit, E. and Kotru, K. and Lauigan, J. and Marjanovic, J. and Meredith, M. and McDonald, M. and Morshead, R. and Narayanaswami, S. and Pawlak, K. A. and Pudenz, K. L. and P\'erez, D. Rodr\'{\i}guez and Sabharwal, P. and Simon, J. and Smull, A. and Sorensen, M. and Stack, D. T. and Stone, M. and Taneja, L. and van de Veerdonk, R. J. M. and Vendeiro, Z. and Weverka, R. T. and White, K. and Wu, T.-Y. and Xie, X. and Zalys-Geller, E. and Zhang, X. and King, J. and Bloom, B. J. and Norcia, M. A.},
  journal = {Phys. Rev. X},
  volume = {15},
  issue = {4},
  pages = {041040},
  numpages = {16},
  year = {2025},
  month = {Dec},
  publisher = {American Physical Society},
  doi = {10.1103/v7ny-fg31},
  url = {https://link.aps.org/doi/10.1103/v7ny-fg31}
}

@article{circularRydberg_Nguyen2018,
  title = {Towards Quantum Simulation with Circular Rydberg Atoms},
  author = {Nguyen, T. L. and Raimond, J. M. and Sayrin, C. and Corti\~nas, R. and Cantat-Moltrecht, T. and Assemat, F. and Dotsenko, I. and Gleyzes, S. and Haroche, S. and Roux, G. and Jolicoeur, Th. and Brune, M.},
  journal = {Phys. Rev. X},
  volume = {8},
  issue = {1},
  pages = {011032},
  numpages = {27},
  year = {2018},
  month = {Feb},
  publisher = {American Physical Society},
  doi = {10.1103/PhysRevX.8.011032},
  url = {https://link.aps.org/doi/10.1103/PhysRevX.8.011032}
}

@article{cantat2020long,
  title = {Long-lived circular Rydberg states of laser-cooled rubidium atoms in a cryostat},
  author = {Cantat-Moltrecht, T. and Corti\~nas, R. and Ravon, B. and M\'ehaignerie, P. and Haroche, S. and Raimond, J. M. and Favier, M. and Brune, M. and Sayrin, C.},
  journal = {Phys. Rev. Res.},
  volume = {2},
  issue = {2},
  pages = {022032},
  numpages = {6},
  year = {2020},
  month = {May},
  publisher = {American Physical Society},
  doi = {10.1103/PhysRevResearch.2.022032},
  url = {https://link.aps.org/doi/10.1103/PhysRevResearch.2.022032}
}

@article{circularRydberg_Cohen2021,
  title = {Quantum Computing with Circular Rydberg Atoms},
  author = {Cohen, Sam R. and Thompson, Jeff D.},
  journal = {PRX Quantum},
  volume = {2},
  issue = {3},
  pages = {030322},
  numpages = {26},
  year = {2021},
  month = {Aug},
  publisher = {American Physical Society},
  doi = {10.1103/PRXQuantum.2.030322},
  url = {https://link.aps.org/doi/10.1103/PRXQuantum.2.030322}
}

@article{circularRydberg_Pultinevicius2025,
author={Pultinevicius, Einius and G{\"o}tzelmann, Aaron and Thielemann, Fabian and H{\"o}lzl, Christian and Meinert, Florian},
  journal={arXiv:2510.27471},
  year={2025},
url={https://arxiv.org/abs/2510.27471}
}

@article{Circular_Sr_Holzl2024,
  title = {Long-Lived Circular Rydberg Qubits of Alkaline-Earth Atoms in Optical Tweezers},
  author = {H\"olzl, C. and G\"otzelmann, A. and Pultinevicius, E. and Wirth, M. and Meinert, F.},
  journal = {Phys. Rev. X},
  volume = {14},
  issue = {2},
  pages = {021024},
  numpages = {11},
  year = {2024},
  month = {May},
  publisher = {American Physical Society},
  doi = {10.1103/PhysRevX.14.021024},
  url = {https://link.aps.org/doi/10.1103/PhysRevX.14.021024}
}

@article{circular-Rydberg_Wu2023,
  title = {Millisecond-Lived Circular Rydberg Atoms in a Room-Temperature Experiment},
  author = {Wu, H. and Richaud, R. and Raimond, J.-M. and Brune, M. and Gleyzes, S.},
  journal = {Phys. Rev. Lett.},
  volume = {130},
  issue = {2},
  pages = {023202},
  numpages = {5},
  year = {2023},
  month = {Jan},
  publisher = {American Physical Society},
  doi = {10.1103/PhysRevLett.130.023202},
  url = {https://link.aps.org/doi/10.1103/PhysRevLett.130.023202}
}

@article{coherent-transport_bluvstein2022,
	title = {A quantum processor based on coherent transport of entangled atom arrays},
	volume = {604},
	copyright = {2022 The Author(s)},
	issn = {1476-4687},
	url = {https://www.nature.com/articles/s41586-022-04592-6},
	doi = {10.1038/s41586-022-04592-6},
	urldate = {2022-10-18},
	journal = {Nature},
	publisher = {Nature Publishing Group},
	author = {Bluvstein, Dolev and Levine, Harry and Semeghini, Giulia and Wang, Tout T. and Ebadi, Sepehr and Kalinowski, Marcin and Keesling, Alexander and Maskara, Nishad and Pichler, Hannes and Greiner, Markus and Vuletić, Vladan and Lukin, Mikhail D.},
	month = apr,
	year = {2022},
	pages = {451--456},
}

@article{99.3_Radnaev2025,
  title = {Universal Neutral-Atom Quantum Computer with Individual Optical Addressing and Nondestructive Readout},
  author = {Radnaev, A.G. and Chung, W.C. and Cole, D.C. and Mason, D. and Ballance, T.G. and Bedalov, M.J. and Belknap, D.A. and Berman, M.R. and Blakely, M. and Bloomfield, I.L. and Buttler, P.D. and Campbell, C. and Chopinaud, A. and Copenhaver, E. and Dawes, M.K. and Eubanks, S.Y. and Friss, A.J. and Garcia, D.M. and Gilbert, J. and Gillette, M. and Goiporia, P. and Gokhale, P. and Goldwin, J. and Goodwin, D. and Graham, T.M. and Guttormsson, C.J. and Hickman, G.T. and Hurtley, L. and Iliev, M. and Jones, E.B. and Jones, R.A. and Kuper, K.W. and Lewis, T.B. and Lichtman, M.T. and Majdeteimouri, F. and Mason, J.J. and McMaster, J.K. and Miles, J.A. and Mitchell, P.T. and Murphree, J.D. and Neff-Mallon, N.A. and Oh, T. and Omole, V. and Parlo Simon, C. and Pederson, N. and Perlin, M.A. and Reiter, A. and Rines, R. and Romlow, P. and Scott, A.M. and Stiefvater, D. and Tanner, J.R. and Tucker, A.K. and Vinogradov, I.V. and Warter, M.L. and Yeo, M. and Saffman, M. and Noel, T.W.},
  journal = {PRX Quantum},
  volume = {6},
  issue = {3},
  pages = {030334},
  numpages = {20},
  year = {2025},
  month = {Aug},
  publisher = {American Physical Society},
  doi = {10.1103/66s8-jj18},
  url = {https://link.aps.org/doi/10.1103/66s8-jj18}
}

@article{99.4_Michael2025,
  title = {Spectroscopy and Modeling of $^{171}\mathrm{Yb}$ Rydberg States for High-Fidelity Two-Qubit Gates},
  author = {Peper, Michael and Li, Yiyi and Knapp, Daniel Y. and Bileska, Mila and Ma, Shuo and Liu, Genyue and Peng, Pai and Zhang, Bichen and Horvath, Sebastian P. and Burgers, Alex P. and Thompson, Jeff D.},
  journal = {Phys. Rev. X},
  volume = {15},
  issue = {1},
  pages = {011009},
  numpages = {30},
  year = {2025},
  month = {Jan},
  publisher = {American Physical Society},
  doi = {10.1103/PhysRevX.15.011009},
  url = {https://link.aps.org/doi/10.1103/PhysRevX.15.011009}
}

@article{erasure_Zhang2025,
      author={Bichen Zhang and Genyue Liu and Guillaume Bornet and Sebastian P. Horvath and Pai Peng and Shuo Ma and Shilin Huang and Shruti Puri and Jeff D. Thompson},
      year={2025},
      journal={arXiv:2506.13724},
      url={https://arxiv.org/abs/2506.13724}
}

@article{99.7_Tsai2025,
  title = {Benchmarking and Fidelity Response Theory of High-Fidelity Rydberg Entangling Gates},
  author = {Tsai, Richard Bing-Shiun and Sun, Xiangkai and Shaw, Adam L. and Finkelstein, Ran and Endres, Manuel},
  journal = {PRX Quantum},
  volume = {6},
  issue = {1},
  pages = {010331},
  numpages = {28},
  year = {2025},
  month = {Feb},
  publisher = {American Physical Society},
  doi = {10.1103/PRXQuantum.6.010331},
  url = {https://link.aps.org/doi/10.1103/PRXQuantum.6.010331}
}

@article{logical_Bluvstein2024,
	title = {Logical quantum processor based on reconfigurable atom arrays},
	volume = {626},
	copyright = {2023 The Author(s)},
	issn = {1476-4687},
	url = {https://www.nature.com/articles/s41586-023-06927-3},
	doi = {10.1038/s41586-023-06927-3},
	number = {7997},
	urldate = {2026-01-15},
	journal = {Nature},
	publisher = {Nature Publishing Group},
	author = {Bluvstein, Dolev and Evered, Simon J. and Geim, Alexandra A. and Li, Sophie H. and Zhou, Hengyun and Manovitz, Tom and Ebadi, Sepehr and Cain, Madelyn and Kalinowski, Marcin and Hangleiter, Dominik and Bonilla Ataides, J. Pablo and Maskara, Nishad and Cong, Iris and Gao, Xun and Sales Rodriguez, Pedro and Karolyshyn, Thomas and Semeghini, Giulia and Gullans, Michael J. and Greiner, Markus and Vuletić, Vladan and Lukin, Mikhail D.},
	month = feb,
	year = {2024},
	pages = {58--65},
}

@article{fault-tolerant_Bluvstein2026,
	title = {A fault-tolerant neutral-atom architecture for universal quantum computation},
	volume = {649},
	copyright = {2025 The Author(s)},
	issn = {1476-4687},
	url = {https://www.nature.com/articles/s41586-025-09848-5},
	doi = {10.1038/s41586-025-09848-5},
	number = {8095},
	urldate = {2026-01-14},
	journal = {Nature},
	publisher = {Nature Publishing Group},
	author = {Bluvstein, Dolev and Geim, Alexandra A. and Li, Sophie H. and Evered, Simon J. and Bonilla Ataides, J. Pablo and Baranes, Gefen and Gu, Andi and Manovitz, Tom and Xu, Muqing and Kalinowski, Marcin and Majidy, Shayan and Kokail, Christian and Maskara, Nishad and Trapp, Elias C. and Stewart, Luke M. and Hollerith, Simon and Zhou, Hengyun and Gullans, Michael J. and Yelin, Susanne F. and Greiner, Markus and Vuletić, Vladan and Cain, Madelyn and Lukin, Mikhail D.},
	month = jan,
	year = {2026},
	pages = {39--46},
}

@article{Dressing_Zeiher2016,
	title = {Many-body interferometry of a {Rydberg}-dressed spin lattice},
	volume = {12},
	copyright = {2016 Springer Nature Limited},
	issn = {1745-2481},
	url = {https://www.nature.com/articles/nphys3835},
	doi = {10.1038/nphys3835},
	number = {12},
	urldate = {2026-01-14},
	journal = {Nature Physics},
	publisher = {Nature Publishing Group},
	author = {Zeiher, Johannes and van Bijnen, Rick and Schauß, Peter and Hild, Sebastian and Choi, Jae-yoon and Pohl, Thomas and Bloch, Immanuel and Gross, Christian},
	month = dec,
	year = {2016},
	pages = {1095--1099},
}

@article{avalanche_Boulier2017,
  title = {Spontaneous avalanche dephasing in large Rydberg ensembles},
  author = {Boulier, T. and Magnan, E. and Bracamontes, C. and Maslek, J. and Goldschmidt, E. A. and Young, J. T. and Gorshkov, A. V. and Rolston, S. L. and Porto, J. V.},
  journal = {Phys. Rev. A},
  volume = {96},
  issue = {5},
  pages = {053409},
  numpages = {11},
  year = {2017},
  month = {Nov},
  publisher = {American Physical Society},
  doi = {10.1103/PhysRevA.96.053409},
  url = {https://link.aps.org/doi/10.1103/PhysRevA.96.053409}
}

@article{Broadening_Goldschmidt2016,
  title = {Anomalous Broadening in Driven Dissipative Rydberg Systems},
  author = {Goldschmidt, E. A. and Boulier, T. and Brown, R. C. and Koller, S. B. and Young, J. T. and Gorshkov, A. V. and Rolston, S. L. and Porto, J. V.},
  journal = {Phys. Rev. Lett.},
  volume = {116},
  issue = {11},
  pages = {113001},
  numpages = {5},
  year = {2016},
  month = {Mar},
  publisher = {American Physical Society},
  doi = {10.1103/PhysRevLett.116.113001},
  url = {https://link.aps.org/doi/10.1103/PhysRevLett.116.113001}
}

@article{continuous_Chiu2025,
	title = {Continuous operation of a coherent 3,000-qubit system},
	volume = {646},
	copyright = {2025 The Author(s)},
	issn = {1476-4687},
	url = {https://www.nature.com/articles/s41586-025-09596-6},
	doi = {10.1038/s41586-025-09596-6},
	urldate = {2026-01-14},
	journal = {Nature},
	publisher = {Nature Publishing Group},
	author = {Chiu, Neng-Chun and Trapp, Elias C. and Guo, Jinen and Abobeih, Mohamed H. and Stewart, Luke M. and Hollerith, Simon and Stroganov, Pavel L. and Kalinowski, Marcin and Geim, Alexandra A. and Evered, Simon J. and Li, Sophie H. and Lyu, Xingjian and Peters, Lisa M. and Bluvstein, Dolev and Wang, Tout T. and Greiner, Markus and Vuletić, Vladan and Lukin, Mikhail D.},
	month = oct,
	year = {2025},
	pages = {1075--1080},
}

@article{continuous_Li2025,
      author={Yiyi Li and Yicheng Bao and Michael Peper and Chenyuan Li and Jeff D. Thompson},
      year={2025},
      journal={arXiv:2506.15633},
      archivePrefix={arXiv},
      url={https://arxiv.org/abs/2506.15633}, 
}

@article{Cong2022,
  title = {Hardware-Efficient, Fault-Tolerant Quantum Computation with Rydberg Atoms},
  author = {Cong, Iris and Levine, Harry and Keesling, Alexander and Bluvstein, Dolev and Wang, Sheng-Tao and Lukin, Mikhail D.},
  journal = {Phys. Rev. X},
  volume = {12},
  issue = {2},
  pages = {021049},
  numpages = {31},
  year = {2022},
  month = {Jun},
  publisher = {American Physical Society},
  doi = {10.1103/PhysRevX.12.021049},
  url = {https://link.aps.org/doi/10.1103/PhysRevX.12.021049}
}

@article{qec_Knill1997,
  title = {Theory of quantum error-correcting codes},
  author = {Knill, Emanuel and Laflamme, Raymond},
  journal = {Phys. Rev. A},
  volume = {55},
  issue = {2},
  pages = {900--911},
  numpages = {0},
  year = {1997},
  month = {Feb},
  publisher = {American Physical Society},
  doi = {10.1103/PhysRevA.55.900},
  url = {https://link.aps.org/doi/10.1103/PhysRevA.55.900}
}

@book{css_code_Gottesman1997,
  title={Stabilizer Codes and Quantum Error Correction},
  author={Gottesman, Daniel},
  year={1997},
  publisher={California Institute of Technology}
}

@article{qec_Shor1995,
  title = {Scheme for reducing decoherence in quantum computer memory},
  author = {Shor, Peter W.},
  journal = {Phys. Rev. A},
  volume = {52},
  issue = {4},
  pages = {R2493--R2496},
  numpages = {0},
  year = {1995},
  month = {Oct},
  publisher = {American Physical Society},
  doi = {10.1103/PhysRevA.52.R2493},
  url = {https://link.aps.org/doi/10.1103/PhysRevA.52.R2493}
}

@article{Surface_code_Fowler2012,
  title = {Surface codes: Towards practical large-scale quantum computation},
  author = {Fowler, Austin G. and Mariantoni, Matteo and Martinis, John M. and Cleland, Andrew N.},
  journal = {Phys. Rev. A},
  volume = {86},
  issue = {3},
  pages = {032324},
  numpages = {48},
  year = {2012},
  month = {Sep},
  publisher = {American Physical Society},
  doi = {10.1103/PhysRevA.86.032324},
  url = {https://link.aps.org/doi/10.1103/PhysRevA.86.032324}
}

@article{99.5-fidelity_Evered2023,
	title = {High-fidelity parallel entangling gates on a neutral-atom quantum computer},
	volume = {622},
	issn = {1476-4687},
	url = {https://www.nature.com/articles/s41586-023-06481-y},
	doi = {10.1038/s41586-023-06481-y},
	number = {7982},
	journal = {Nature},
	publisher = {Nature Publishing Group},
	author = {Evered, Simon J. and Bluvstein, Dolev and Kalinowski, Marcin and Ebadi, Sepehr and Manovitz, Tom and Zhou, Hengyun and Li, Sophie H. and Geim, Alexandra A. and Wang, Tout T. and Maskara, Nishad and Levine, Harry and Semeghini, Giulia and Greiner, Markus and Vuletić, Vladan and Lukin, Mikhail D.},
	month = oct,
	year = {2023},
	pages = {268--272},
}

@article{arc_sibalic2017,
	title = {{ARC}: {An} open-source library for calculating properties of alkali {Rydberg} atoms},
	volume = {220},
	issn = {0010-4655},
	shorttitle = {{ARC}},
	url = {https://www.sciencedirect.com/science/article/pii/S0010465517301972},
	doi = {10.1016/j.cpc.2017.06.015},
	journal = {Comput. Phys. Commun.},
	author = {Šibalić, N. and Pritchard, J. D. and Adams, C. S. and Weatherill, K. J.},
	month = nov,
	year = {2017},
	pages = {319--331},
}

@article{6100tweezer_manetsch2025,
	title = {A tweezer array with 6,100 highly coherent atomic qubits},
	volume = {647},
	copyright = {2025 The Author(s)},
	issn = {1476-4687},
	url = {https://www.nature.com/articles/s41586-025-09641-4},
	doi = {10.1038/s41586-025-09641-4},
	
	number = {8088},
	journal = {Nature},
	publisher = {Nature Publishing Group},
	author = {Manetsch, Hannah J. and Nomura, Gyohei and Bataille, Elie and Lv, Xudong and Leung, Kon H. and Endres, Manuel},
	month = nov,
	year = {2025},
	pages = {60--67},
}

@article{2000rearrange_Pichard2024,
  title = {Rearrangement of individual atoms in a 2000-site optical-tweezer array at cryogenic temperatures},
  author = {Pichard, Gr\'egoire and Lim, Desiree and Bloch, \'Etienne and Vaneecloo, Julien and Bourachot, Lilian and Both, Gert-Jan and M\'eriaux, Guillaume and Dutartre, Sylvain and Hostein, Richard and Paris, Julien and Ximenez, Bruno and Signoles, Adrien and Browaeys, Antoine and Lahaye, Thierry and Dreon, Davide},
  journal = {Phys. Rev. Appl.},
  volume = {22},
  issue = {2},
  pages = {024073},
  numpages = {7},
  year = {2024},
  month = {Aug},
  publisher = {American Physical Society},
  doi = {10.1103/PhysRevApplied.22.024073},
  url = {https://link.aps.org/doi/10.1103/PhysRevApplied.22.024073}
}

@article{6000s_Schymik2021,
  title = {Single Atoms with 6000-Second Trapping Lifetimes in Optical-Tweezer Arrays at Cryogenic Temperatures},
  author = {Schymik, Kai-Niklas and Pancaldi, Sara and Nogrette, Florence and Barredo, Daniel and Paris, Julien and Browaeys, Antoine and Lahaye, Thierry},
  journal = {Phys. Rev. Appl.},
  volume = {16},
  issue = {3},
  pages = {034013},
  numpages = {8},
  year = {2021},
  month = {Sep},
  publisher = {American Physical Society},
  doi = {10.1103/PhysRevApplied.16.034013},
  url = {https://link.aps.org/doi/10.1103/PhysRevApplied.16.034013}
}

@article{3000s_Zhang2025,
  title = {High Optical Access Cryogenic System for Rydberg Atom Arrays with a 3000-Second Trap Lifetime},
  author = {Zhang, Zhenpu and Hsu, Ting-Wei and Tan, Ting You and Slichter, Daniel H. and Kaufman, Adam M. and Marinelli, Matteo and Regal, Cindy A.},
  journal = {PRX Quantum},
  volume = {6},
  issue = {2},
  pages = {020337},
  numpages = {20},
  year = {2025},
  month = {May},
  publisher = {American Physical Society},
  doi = {10.1103/PRXQuantum.6.020337},
  url = {https://link.aps.org/doi/10.1103/PRXQuantum.6.020337}
}

@article{ITO_Meinert2020,
  title = {Indium tin oxide films meet circular Rydberg atoms: Prospects for novel quantum simulation schemes},
  author = {Meinert, Florian and H\"olzl, Christian and Nebioglu, Mehmet Ali and D'Arnese, Alessandro and Karl, Philipp and Dressel, Martin and Scheffler, Marc},
  journal = {Phys. Rev. Res.},
  volume = {2},
  issue = {2},
  pages = {023192},
  numpages = {7},
  year = {2020},
  month = {May},
  publisher = {American Physical Society},
  doi = {10.1103/PhysRevResearch.2.023192},
  url = {https://link.aps.org/doi/10.1103/PhysRevResearch.2.023192}
}

@Article{Cao2024DressingMetrology,
author={Cao, Alec
and Eckner, William J.
and Lukin Yelin, Theodor
and Young, Aaron W.
and Jandura, Sven
and Yan, Lingfeng
and Kim, Kyungtae
and Pupillo, Guido
and Ye, Jun
and Darkwah Oppong, Nelson
and Kaufman, Adam M.},
title={Multi-qubit gates and Schr{\"o}dinger cat states in an optical clock},
journal={Nature},
year={2024},
month={Oct},
day={01},
volume={634},
number={8033},
pages={315-320},
issn={1476-4687},
doi={10.1038/s41586-024-07913-z},
url={https://doi.org/10.1038/s41586-024-07913-z}
}

@article{Schleier-Smith2023DressingSqueezing,
  title = {Spin Squeezing by Rydberg Dressing in an Array of Atomic Ensembles},
  author = {Hines, Jacob A. and Rajagopal, Shankari V. and Moreau, Gabriel L. and Wahrman, Michael D. and Lewis, Neomi A. and Markovi\ifmmode \acute{c}\else \'{c}\fi{}, Ognjen and Schleier-Smith, Monika},
  journal = {Phys. Rev. Lett.},
  volume = {131},
  issue = {6},
  pages = {063401},
  numpages = {7},
  year = {2023},
  month = {Aug},
  publisher = {American Physical Society},
  doi = {10.1103/PhysRevLett.131.063401},
  url = {https://link.aps.org/doi/10.1103/PhysRevLett.131.063401}
}

@article{Spencer1982Na,
  title = {Temperature dependence of blackbody-radiation-induced transfer among highly excited states of sodium},
  author = {Spencer, William P. and Vaidyanathan, A. Ganesh and Kleppner, Daniel and Ducas, Theodore W.},
  journal = {Phys. Rev. A},
  volume = {25},
  issue = {1},
  pages = {380--384},
  numpages = {0},
  year = {1982},
  month = {Jan},
  publisher = {American Physical Society},
  doi = {10.1103/PhysRevA.25.380},
  url = {https://link.aps.org/doi/10.1103/PhysRevA.25.380}
}

@article{Spencer1981Na,
  title = {Measurements of lifetimes of sodium Rydberg states in a cooled environment},
  author = {Spencer, William P. and Vaidyanathan, A. Ganesh and Kleppner, Daniel and Ducas, Theodore W.},
  journal = {Phys. Rev. A},
  volume = {24},
  issue = {5},
  pages = {2513--2517},
  numpages = {0},
  year = {1981},
  month = {Nov},
  publisher = {American Physical Society},
  doi = {10.1103/PhysRevA.24.2513},
  url = {https://link.aps.org/doi/10.1103/PhysRevA.24.2513}
}

@article{Gallagher1979BBRTheory,
  title = {Interactions of Blackbody Radiation with Atoms},
  author = {Gallagher, T. F. and Cooke, W. E.},
  journal = {Phys. Rev. Lett.},
  volume = {42},
  issue = {13},
  pages = {835--839},
  numpages = {0},
  year = {1979},
  month = {Mar},
  publisher = {American Physical Society},
  doi = {10.1103/PhysRevLett.42.835},
  url = {https://link.aps.org/doi/10.1103/PhysRevLett.42.835}
}

@article{AtomComputing2025Gate,
  title = {High-Fidelity Universal Gates in the ${}^{171}$$\mathrm{Yb}$ Ground-State Nuclear-Spin Qubit},
  author = {Muniz, J. A. and Stone, M. and Stack, D. T. and Jaffe, M. and Kindem, J. M. and Wadleigh, L. and Zalys-Geller, E. and Zhang, X. and Chen, C.-A. and Norcia, M. A. and Epstein, J. and Halperin, E. and Hummel, F. and Wilkason, T. and Li, M. and Barnes, K. and Battaglino, P. and Bohdanowicz, T. C. and Booth, G. and Brown, A. and Brown, M. O. and Cairncross, W. B. and Cassella, K. and Coxe, R. and Crow, D. and Feldkamp, M. and Griger, C. and Heinz, A. and Jones, A. M. W. and Kim, H. and King, J. and Kotru, K. and Lauigan, J. and Marjanovic, J. and Megidish, E. and Meredith, M. and McDonald, M. and Morshead, R. and Narayanaswami, S. and Nishiguchi, C. and Paule, T. and Pawlak, K. A. and Pudenz, K. L. and P\'erez, D. Rodr\'{\i}guez and Ryou, A. and Simon, J. and Smull, A. and Urbanek, M. and van de Veerdonk, R. J. M. and Vendeiro, Z. and Wu, T.-Y. and Xie, X. and Bloom, B. J.},
  journal = {PRX Quantum},
  volume = {6},
  issue = {2},
  pages = {020334},
  numpages = {18},
  year = {2025},
  month = {May},
  publisher = {American Physical Society},
  doi = {10.1103/PRXQuantum.6.020334},
  url = {https://link.aps.org/doi/10.1103/PRXQuantum.6.020334}
}
